\documentclass[osajnl,twocolumn,showpacs,superscriptaddress,10pt]{revtex4-1} 
\usepackage{amsmath,amssymb,graphicx}
\usepackage[breaklinks]{hyperref}

\begin{document}

\title{Alignment sensing and control for \\ squeezed vacuum states of light}

\author{E. Schreiber}
\email{Corresponding author: emil.schreiber@aei.mpg.de}
\affiliation{Max-Planck-Institut f\"ur Gravitationsphysik (Albert-Einstein-Institut) and Leibniz Universit\"at Hannover, Callinstr.\,38, 30167 Hannover, Germany}

\author{K.\,L. Dooley}
\thanks{Current address: The University of Mississippi, University, MS38677, USA}
\affiliation{Max-Planck-Institut f\"ur Gravitationsphysik (Albert-Einstein-Institut) and Leibniz Universit\"at Hannover, Callinstr.\,38, 30167 Hannover, Germany}

\author{H. Vahlbruch}
\author{C. Affeldt}
\author{A. Bisht}
\author{J.\,R. Leong}
\author{J. Lough}
\affiliation{Max-Planck-Institut f\"ur Gravitationsphysik (Albert-Einstein-Institut) and Leibniz Universit\"at Hannover, Callinstr.\,38, 30167 Hannover, Germany}

\author{M. Prijatelj}
\affiliation{European Gravitational Observatory (EGO), I-56021 Cascina (Pi), Italy}

\author{J. Slutsky}
\affiliation{CRESST and Gravitational Astrophysics Laboratory NASA/GSFC, Greenbelt, MD~20771, USA}

\author{M. Was}
\affiliation{Laboratoire d'Annecy-le-Vieux de Physique des Particules (LAPP), Universit\'e de Savoie, CNRS/IN2P3, F-74941 Annecy-le-Vieux, France}

\author{H. Wittel}
\author{K. Danzmann}
\author{H. Grote}
\affiliation{Max-Planck-Institut f\"ur Gravitationsphysik (Albert-Einstein-Institut) and Leibniz Universit\"at Hannover, Callinstr.\,38, 30167 Hannover, Germany}

\begin{abstract}
Beam alignment is an important practical aspect of the application of squeezed states of light.
Misalignments in the detection of squeezed light result in a reduction of the observable squeezing level.
In the case of squeezed vacuum fields that contain only very few photons, special measures must be taken in order to sense and control the alignment of the essentially dark beam.
The GEO\,600 gravitational wave detector employs a squeezed vacuum source to improve its detection sensitivity beyond the limits set by classical quantum shot noise. 
Here, we present our design and implementation of an alignment sensing and control scheme that ensures continuous optimal alignment of the squeezed vacuum field at GEO\,600 on long time scales in the presence of free-swinging optics. 
This first demonstration of a squeezed light automatic alignment system will be of particular interest for future long-term applications of squeezed vacuum states of light.
\end{abstract}


\maketitle 

One of the first practical applications of squeezed states of light is in interferometric gravitational wave detection.
Injecting a squeezed vacuum state into the dark output port of a Michelson interferometer can reduce the quantum noise in the observed light field quadrature and thereby improve the sensitivity of the detector at frequencies where quantum noise is a limiting factor~\cite{Caves81, Schnabel2010}. This scheme has been successfully demonstrated at the GEO\,600~\cite{SQZGEO11} and LIGO gravitational wave detectors~\cite{SQZLIGO13} and it is in constant use at GEO\,600 since 2011~\cite{Grote13, Affeldt14}.

To observe the full benefit of squeezing, the squeezed light field must be well overlapped with the interferometer's output beam.
The two fields have to match in their alignment axes, beam parameters, and relative phase.
While phase control has been treated in~\cite{Dooley15}, this letter reports on the first design and implementation of an alignment sensing and control scheme for the squeezed vacuum field in the context of GEO\,600.

\begin{figure}[ht]
\centerline{\includegraphics[width=1\columnwidth]{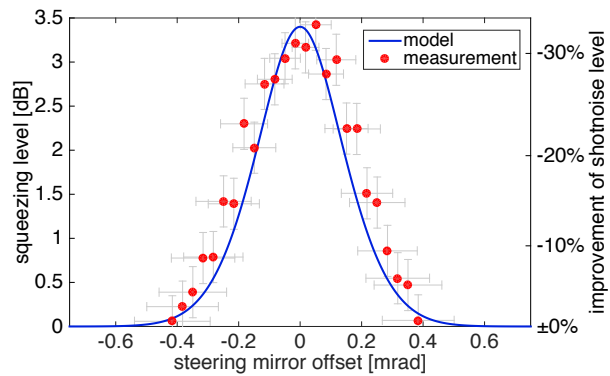}}
\caption{
The effect of misalignment on the observed squeezing level. Here the squeezed beam is misaligned by rotating one of the mirrors in the input path and the squeezing level is recorded. With $0.4\,\mathrm{mrad}$ of misalignment, nearly all squeezing is lost.
The measurements are compared to a numerical model~\cite{note:model}.
The small remaining difference along the x-axis between measurements and model is consistent with the uncertainty of the actuator calibration.
}
\label{fig:misalignment}
\end{figure}

Misalignments need to be compensated because a reduced overlap of the squeezed field with the interferometer beam will lead to effective optical losses~\cite{Oelker14}, thus degrading the observed squeezing level and the corresponding improvement of sensitivity (see Figure~\ref{fig:misalignment}).
Additionally, alignment fluctuations can introduce lock-point errors in the phase control~\cite{Oelker14, Dooley15}.

While short-term tabletop experiments are typically stable enough that one-off manual alignment is sufficient, in a large-scale optical system, such as GEO\,600, drifts of the individual components will occur and have to be actively controlled.
To always maintain optimal alignment of the squeezed light field, we have developed a method to continuously sense and automatically actuate on the alignment of the squeezed beam.

\begin{figure}
\centerline{\includegraphics[width=1\columnwidth]{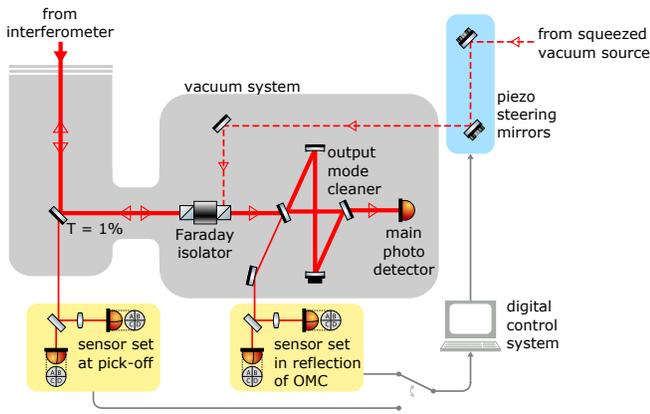}}
\caption{Simplified layout of the GEO\,600 output optics with squeezing injection. The squeezed vacuum field is injected via a Faraday isolator, enters the interferometer through the output port, and is reflected back. It then travels together with the interferometer output beam through the output mode cleaner (OMC) to the detection photodiode. A pair of 3-axis piezo-actuated mirrors in the in-air path serves to steer the squeezed light field. Sets of differential wavefront sensors (DWSs) at two possible locations can be used for alignment sensing.}
\label{fig:layout}
\end{figure}

Figure~\ref{fig:layout} shows the output path of GEO\,600 where the squeezed vacuum field is injected into the interferometer.
The main interferometer optics are housed inside a vacuum envelope and are seismically isolated with individual active multi-stage pendulums for each mirror.
An output mode cleaner (OMC) which serves to eliminate disturbing higher order spatial modes from the interferometer's output beam sits on a passive multi-stage isolation system, also in vacuum.
The squeezed light source is located in air on a passively isolated optical table inside an acoustically shielded box.
These different components that make up the injection path for the squeezed field have no rigid mechanical connection and thus their relative alignment is subject to small but significant drifts in many degrees of freedom (DOFs), only some of which are already controlled by other alignment systems of the interferometer~\cite{Prijatelj10}.

Without control, we observe drifts over a timescale of days that misalign the squeezed beam from the interferometer's output beam, eventually leading to an observable reduction of the squeezing level.
This is especially the case in the presence of environmental disturbances such as small temperature changes or during times of instrumental work on parts of the interferometer that might affect alignments.
Faster alignment fluctuations, such as those caused by residual pendulum movements at the resonance frequencies around $1\,\mathrm{Hz}$, are present but currently not limiting.
This will, however, change with future reductions of other optical injection losses, making the effective losses due to misalignments more important.
It is therefore desirable to have an active control system for the alignment of the squeezed light field that can sense and suppress both long-term drifts and fast fluctuations up to several hertz.

For the control we need alignment error signals that give a measure of how well the axis of the squeezed field overlaps with the interferometer beam axis.
To generate these error signals we use differential wavefront sensing~\cite{Morrison1994}, in which the beat of two light fields with different frequencies is detected on a quadrant photodiode and demodulated to give a signal proportional to the relative misalignments of the two beams~\cite{note:scanners}.
With a set of two differential wavefront sensors (DWSs) at different positions along the beam (with a different Gouy phase), the full set of four alignment DOFs (angular and lateral offsets of the beam axis in horizontal and vertical direction) can be sensed.

Since the squeezed vacuum field contains only very few photons, it cannot directly be detected by the DWSs.
Instead, the axis of the squeezed vacuum beam is marked by coherent control sidebands (CCSBs)~\cite{Vahlb06}.
These are auxiliary light fields with a frequency offset of $\pm15.2\,\mathrm{MHz}$ with respect to the carrier frequency.
They are generated inside the squeezed light source and are also used for the longitudinal phase control of the squeezed vacuum~\cite{Dooley15, Vahlbruch10}.
The CCSBs resonate inside the cavity of the optical parametric amplifier (OPA).
This ensures that they share the same spatial mode as the squeezed light field that originates from within the OPA.

\begin{figure}
\centerline{\includegraphics[width=1\columnwidth]{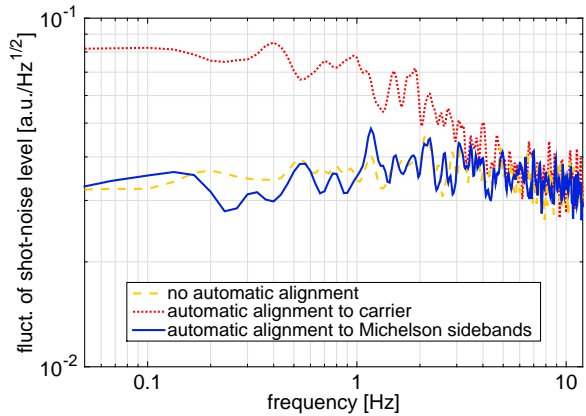}}
\caption{Squeezing level stability with different alignment signals.
The plot shows spectra of the detector's shot-noise level calculated as a band-limited RMS (BLRMS) in the frequency band from $4$ to $5\,\mathrm{kHz}$.
When locking the alignment of the squeezed beam to the carrier, the contamination of HOMs leads to excess fluctuations within the bandwidth of the alignment loops, which was $4\,\mathrm{Hz}$ in this example.
When locking to the Michelson sidebands instead, no degradation occurs.
There is no measurable improvement with respect to the case without any alignment actuation because alignment fluctuations beyond very slow drifts are currently not a limiting source of squeezing losses.
}
\label{fig:CarrierVsMISB}
\end{figure}

We can now either detect the beat of the CCSBs with the interferometer's carrier light, or alternatively with existing control sidebands at $\pm14.9\,\mathrm{MHz}$ (called Michelson sidebands) co-travelling with the interferometer beam.
The carrier light at the output port, before it is filtered by the OMC, is not a perfect Gaussian beam but is composed largely of higher order modes (HOMs).
The distribution of the HOMs is not static in time and introduces varying offsets in the DWS signals which show up as false misalignment signals.
The Michelson sidebands, on the other hand, are spatially much cleaner~\cite{note:HOM} and thus can provide alignment signals that are less contaminated by HOMs.
Both variants have been implemented at GEO\,600 using the same hardware.
The field used for generating the DWS error signals can be selected by choosing the demodulation frequency for the respective optical beat ($15.2\,\mathrm{MHz}$ for CCSBs vs. carrier light, $0.3\,\mathrm{MHz}$ for CCSBs vs. Michelson sidebands).
As expected, using the Michelson sidebands rather than the carrier yields significantly more accurate alignment signals.
The effect of HOMs in the carrier-derived signals dominates in a way such that using them for feedback control leads to a strongly fluctuating squeezing level as shown in Figure~\ref{fig:CarrierVsMISB}.
No such effects are present when aligning the squeezed field to the Michelson sidebands, so this proved to be the better method and it was adopted for permanent operation.

The DWSs can be placed at two different points in the output chain (see Figure~\ref{fig:layout}):
One of the steering mirrors in front of the OMC has a transmission of $1\,\%$ to provide a pick-off of all relevant fields.
Alternatively, the CCSBs and Michelson sidebands are also available in reflection of the OMC.
The latter variant has the benefit of higher light powers, allowing a better signal-to-noise ratio.
Also, it opens up the possibility of altogether avoiding the loss-inducing pick-off.
Sets of DWSs at both positions have been set up and are currently used interchangeably.

Using the error signals, the aim is to actively control the alignment of the squeezed beam with feedback loops.
A set of two 3-axis piezo-actuated mirrors in the in-air part of the squeezing injection path (see Figure~\ref{fig:layout}) serve as alignment actuators.
They can shift the angle of the beam axis at their respective positions horizontally and vertically and each provide a total range of about $2.5\,\mathrm{mrad}$ in the current setup.
Like the sensors, the two actuators are spaced apart along the beam path at different Gouy phases to ensure accessibility of all four alignment DOFs.

\begin{figure}
\centerline{\includegraphics[width=1\columnwidth]{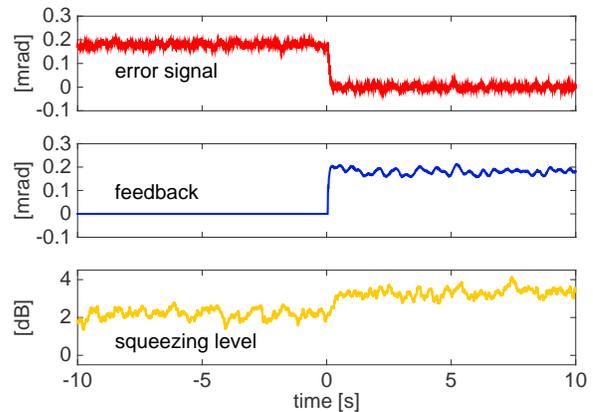}}
\caption{Switching the automatic alignment on.
Starting from a situation with significant misalignment, the alignment loops are activated.
The loop feedbacks compensate the misalignment, driving the error points to zero, and as a result the observed squeezing level increases.
}
\label{fig:AAon}
\end{figure}

\begin{figure}
\centerline{\includegraphics[width=1\columnwidth]{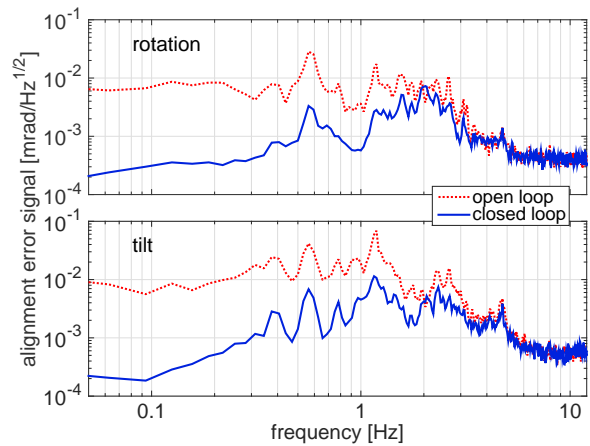}}
\caption{
In-loop suppression of the alignment error signals while running the automatic alignment system for two DOFs.
Here the signals from one DWS at the pick-off are used and fed back to one of the actuators.
The error signals are calibrated to milliradians of angular movement of the steering mirror.
The prominent features in the spectra around $1\,\mathrm{Hz}$ are related to residual pendulum movements and intentional dithers of the interferometer's output optics.
}
\label{fig:EPspec}
\end{figure}

The feedback control loops are implemented in GEO\,600's digital control and data system (CDS).
The demodulated DWS signals are digitized, filtered digitally, and then sent back to analog high-voltage amplifiers that drive the piezo actuators.
Figure~\ref{fig:AAon} is a simple demonstration of the positive effect of the alignment control when starting from a misaligned state.
Figure~\ref{fig:EPspec} shows the resulting error-point suppression of the control loop for two DOFs.
Unity-gain frequencies of up to $4\,\mathrm{Hz}$ are easily achieved.

The alignment control system for the squeezed field has been in operation for most of the time since squeezing was implemented at GEO\,600.
The typical state of operation is to control two DOFs using one DWS and one actuator.
This is currently sufficient for suppressing almost all relevant misalignments over long periods of time.
The remaining two uncontrolled DOFs have to be tuned manually every few months.
Full automatic control of all four DOFs has also been successfully demonstrated and is currently being fully commissioned for future permanent use.

In conclusion, we have designed and implemented a system for sensing and controlling the relative alignment of a squeezed vacuum field to a given interferometer light field.
Since there are virtually no photons in the squeezed vacuum state itself, we have resorted to aligning a sideband field to our interferometer.
The system we have implemented uses digital control loops which are relatively fast, having a unity gain frequency of a few hertz.
It has been running at the gravitational wave detector GEO\,600 over a long period of time, demonstrating its reliability.
Having automatic control of the alignment proved to be a valuable asset for upkeeping optimal squeezing performance, both during commissioning work as well as during science operation of GEO\,600.
Because upcoming applications of squeezed light in gravitational wave interferometers will aim for higher levels of squeezing and have lower optical losses, the demands on alignment will be more stringent than those seen in GEO\,600 at the time of writing this letter.
However, since the control loops we have implemented are fast enough to suppress most occurring alignment fluctuations, they are already well-equipped to fulfill these stronger alignment needs.
Furthermore, as the challenge of achieving good alignment will exist in any long-term squeezing experiments, this system may also find applications outside the realm of gravitational wave interferometry.

\section*{Acknowledgements}
The authors are grateful for support from the Science and Technology Facilities Council (STFC), the University of Glasgow in the UK, the Bundesministerium f\"ur Bildung und Forschung (BMBF), and the state of Lower Saxony in Germany.
This work was partly supported by DFG grant SFB/Transregio 7 Gravitational Wave Astronomy.

\medskip \noindent
This document has been assigned LIGO document number P1500056.


\begin{thebibliography}{99}
\bibitem{note:model}
We used the interferometer simulation software \mbox{\sc Finesse}~\cite{finesse} to calculate for different misalignments the overlap of the squeezed light field with the interferometer beam as given by the OMC's eigenmode. Effects of HOMs can be neglected in this case because they are suppressed by the OMC.
\bibitem{note:scanners}
While differential wavefront sensing ideally gives a purely relative signal, common movements of the two beams on the quadrant photodiode can introduce small errors.
To mitigate this we use active scanners that keep the Michelson sidebands centered on the sensors by detecting the beat of the two sidebands at twice the modulation frequency~\cite{Grote10}.
\bibitem{note:HOM}
The Michelson sidebands at the interferometer output are less contaminated by HOMs than the carrier. Due to their shifted frequency and the interferometer's Schnupp asymmetry they are further away from the destructive interference condition.
Thus more of the fundamental mode leaves the output port, making it the dominant mode contribution.
\end{thebibliography}
\end{document}